# Electron Phase Detection in Single Molecules by Interferometry

Zhixin Chen[1]*, Jie-Ren Deng[2], Mengyun Wang[1], Nikolaos Farmakidis[1], Jonathan Baugh[3], Harish Bhaskaran[1], Jan A. Mol[4], Harry L. Anderson[2], Lapo Bogani[1,5]*, and James O. Thomas[1,4]*

[1]Department of Materials, University of Oxford, Parks Road, Oxford, OX1 3PH, UK

[2]Department of Chemistry, University of Oxford, Chemistry Research Laboratory, Oxford, OX1 3TA, UK

[3]Institute for Quantum Computing, University of Waterloo, Waterloo, ON N2L 3G1, Canada

[4]School of Physical and Chemical Sciences, Queen Mary University of London, London, E1 4NS, UK

[5]Departments of Chemistry and Physics, University of Florence, Sesto Fiorentino, 50019, Italy

*zhixin.chen@materials.ox.ac.uk, *lapo.bogani@unifi.it, *j.o.thomas@qmul.ac.uk

**Interferometry has underpinned a century of discoveries, ranging from the disproval of the ether theory to the detection of gravitational waves, offering insights into wave dynamics with unrivalled precision through the measurement of phase relationships.[1] In electronics, phase-sensitive measurements can probe the nature of transmissive topological and quantum states, but are only possible using complex device structures in magnetic fields.[2-4] Here we demonstrate electronic interferometry in a single-molecule device through the study of non-equilibrium Fano resonances. We show the phase difference between an electronic orbital and a coupled Fabry-Perot resonance are tuneable through electric fields, and consequently it is possible to read out quantum information in the smallest devices, offering new avenues for the coherent manipulation down to single molecules.**

The relative phases acquired by electron waves propagating through a device encode fundamental information about the symmetries of the transmissive states. However, in electronics, phase information is difficult to retrieve through interferometry, because electron wavelengths in metals are typically on the nanometre scale, orders of magnitude smaller than UV-visible optical equivalents. Successful implementations have made use of magnetic-field dependent measurements on device structures with large (micron-sized) footprints, exploiting the Aharonov-Bohm effect to observe phase shifts between geometrically-separated pathways at millikelvin temperature.[5-8] It is not possible to implement this approach for devices on the scale of a few nanometres, due to the large magnetic fields that would be required, and the inability to define two coherent channels at this size lithographically. Effectively, this leaves a blind spot in phase-sensitive measurement techniques for devices that incorporate systems such as molecules, graphene nanoribbons, or few-nm quantum dots, which can display interesting topological properties and are promising for higher temperature operation. New methods for phase-sensitive measurements are required for these devices, where symmetry and topology can dominate their electronic structure, and consequently, device response.[9,10]

Our approach to electronic interferometry bridges meso- and molecular- scales by embedding a single-molecule junction in a graphene Fabry-Pérot (FP) cavity.[11] The two channels that undergo interference are *(i)* an electronic resonance of a single molecular orbital and *(ii)* the coherent FP transmission channels within the graphene. These are coupled to create the electronic equivalent of a resonator-waveguide interferometer (Fig. 1a). Single molecules are useful testbeds as they are atomically defined, giving them known quantum states with phase properties that are readily modelled.[12-14] The detuning of the two channels is precisely controlled with electric fields through their different capacitive couplings to the bias ($V_{sd}$) and gate voltages ($V_g$). The experimental signatures of their interference are Fano resonances, from which transmission phase is measured. In the design, geometrically separated transmission paths are not required, and we do not rely on an external magnetic field to control the phase difference between them.

For operation, the resonant states of the molecule need to support electron-phase coherence over distances larger than the tunnel junction.[15] Our molecules of choice are thus edge-fused porphyrin nanoribbons, consisting of 8 and 18 repeating units (**FP8**, ~8 nm long and **FP18**, ~16 nm long hereafter; see structures in Fig. 3), which support highly delocalised electronic wavefunctions.[16,17] These ribbons also provide the required efficient electronic coupling to the graphene FP cavity by π-stacking interactions (Fig.

1a), with intermediate electronic coupling.[11] The electronic FP cavity in graphene is first created by Joule-heating (see Methods for more details). To accomplish this, we pattern graphene into a bow-tie shaped constriction that is 100 nm wide at the narrowest point (Fig. 1b). Next, we apply voltage ramps across the constriction. The voltage ramps have two effects on the graphene: to electrically break it down at the centre of the constriction into a nanogap tunnel junction (width 1-2 nm),[18,19] and second, to anneal it. Graphene on $HfO_2$ is *p*-doped by the substrate,[20] and the annealing further increases the hole concentration around the constriction, resulting in a highly doped $p^+$ region that is 0.8 – 1.0 μm across, as demonstrated by Kelvin probe force microscopy (in Fig. 1c), forming a potential well that functions as the electronic FP cavity. The molecule, **FP8** or **FP18**, is then drop-cast from dilute solution onto the graphene, bridging the nanogap at the centre of the FP cavity (Supplementary Fig S3-1 and S3-3).

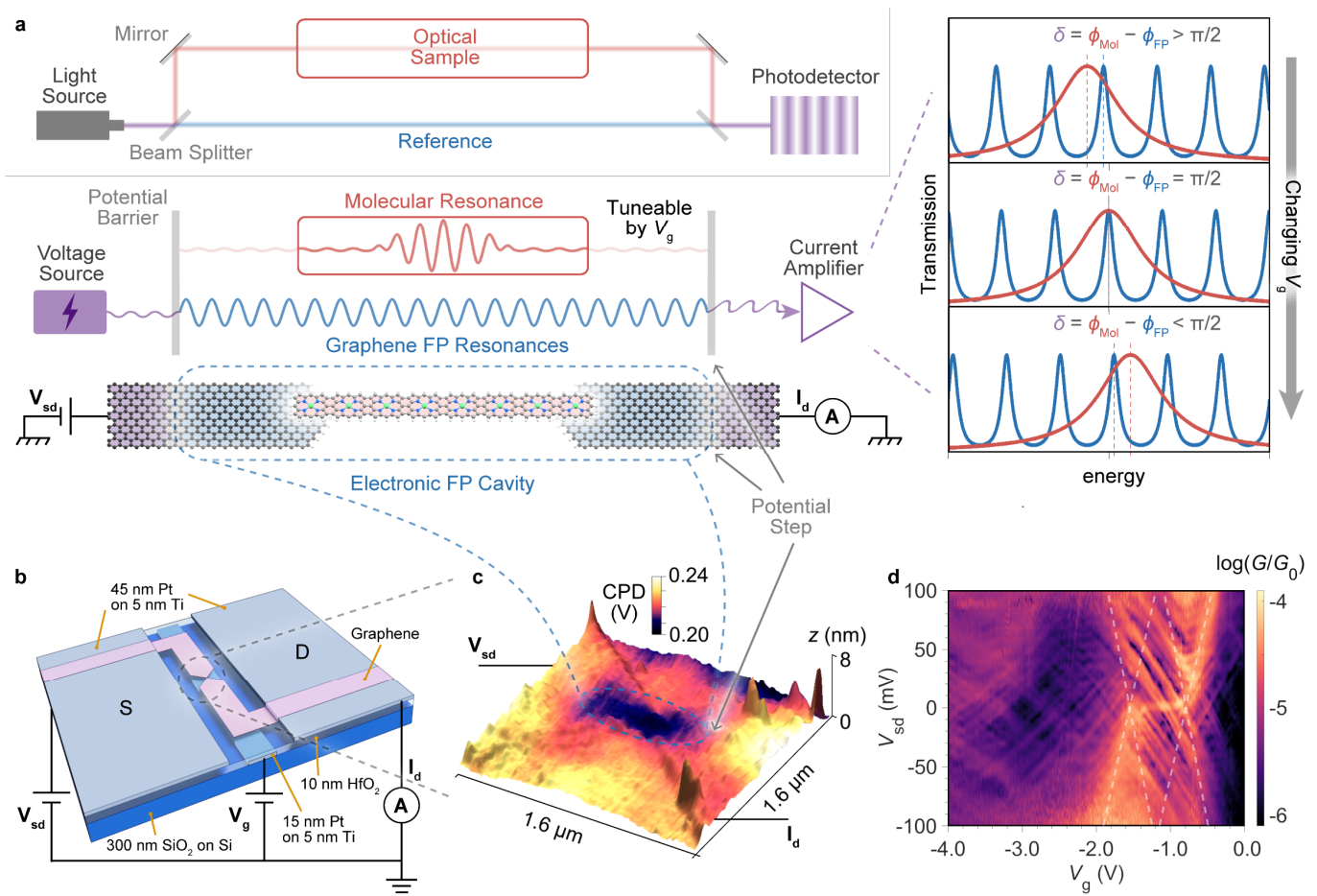

**Fig. 1 | Concept for electron phase detection. a** The comparison of an electron interferometer concept for a single molecule in a graphene-based electron interferometer (not to scale) with an optical scheme. The gate voltage of the electronic device has a different capacitive coupling to the FP and molecular resonances, allowing precise tuning of their relative energies and transmission phase difference ($\delta = \phi_{Mol} - \phi_{FP}$), as shown in the right-hand panels. **b** Device architecture. The grey-blue rectangular strip in the centre is the local platinum gate electrode under a 10 nm layer of $HfO_2$ (transparent); the rectangular areas (grey-blue) at both ends are source and drain platinum electrodes, which are in contact with the bowtie-shape graphene (pink). **c** Contact potential difference (CPD) measurements of a graphene constriction after Joule heating using Kelvin probe force microscopy. The region of graphene annealed during electroburning is ~ 800 – 1000 nm in length, and visible as a $p^+$ doped dark region in the CPD map. **d** The interference fringes in differential conductance ($G = dI_{sd}/dV_{sd}$) map measured as a function of bias voltage ($V_{sd}$) and gate voltage ($V_g$). The white dashed lines outline molecular resonances. The conductance is plotted in logarithmic scale as the ratio to conductance quantum $G_0 = 2e^2/h$ where *e* is the elementary charge and *h* is Planck's constant.

The phase information is obtainable from the three-terminal operation of the device by measuring the differential conductance $G = dI_{sd}/dV_{sd}$ as a function of bias ($V_{sd}$) and gate ($V_g$) voltages. The resulting conductance map shows Coulomb blockade diamonds that arise from resonant molecular transport superimposed on interference fringes of the FP cavity. Throughout the map, electron transport through the device can only occur via the porphyrin nanoribbon, however, within the Coulomb diamonds it is off-resonant (but elastic/phase-coherent) with respect to the molecular orbitals, with the conductance peaking when the FP resonance condition is met. Interestingly, the conductance of the FP modes depend on the charge state ($N$) of the nanoribbon (Fig 1b), with $N$ the number of electrons on the neutral molecule. Enhanced conductance throughout the $N-1$ charge state region (Fig. 1d, 2a, 2b), is in line with EPR data that shows polarons generated by oxidation are coherently delocalised over the porphyrin nanoribbon length (10–14 porphyrin units),[17] and reports from STM break junction measurements that conductance increases,[21] and ballistic transport can be supported,[22] when the molecules are oxidised to a radial cation ($N-1$). Furthermore, there is a Kondo peak through $N-1$ which gives an additional conductance increase, but only at zero bias voltage. Overall, the transport data confirm the coexistence of the FP resonances and molecular orbital resonances, as required for the phase detection scheme in Fig. 1a. Furthermore, the slopes of the two channels shows that the back gate, buried under 10 nm of $HfO_2$ dielectric, has different couplings, $\alpha_i (= C_i/C_{tot}$, where $C_i$ is the capacitance of channel $i$ to the gate electrode, and $C_{tot}$ is the sum of device capacitances) to them,[23] with $\alpha_{FP} = 0.05$ and $\alpha_{Mol} = 0.22$, so their energy difference is continuously tuneable with voltage (Fig. 1a panels).

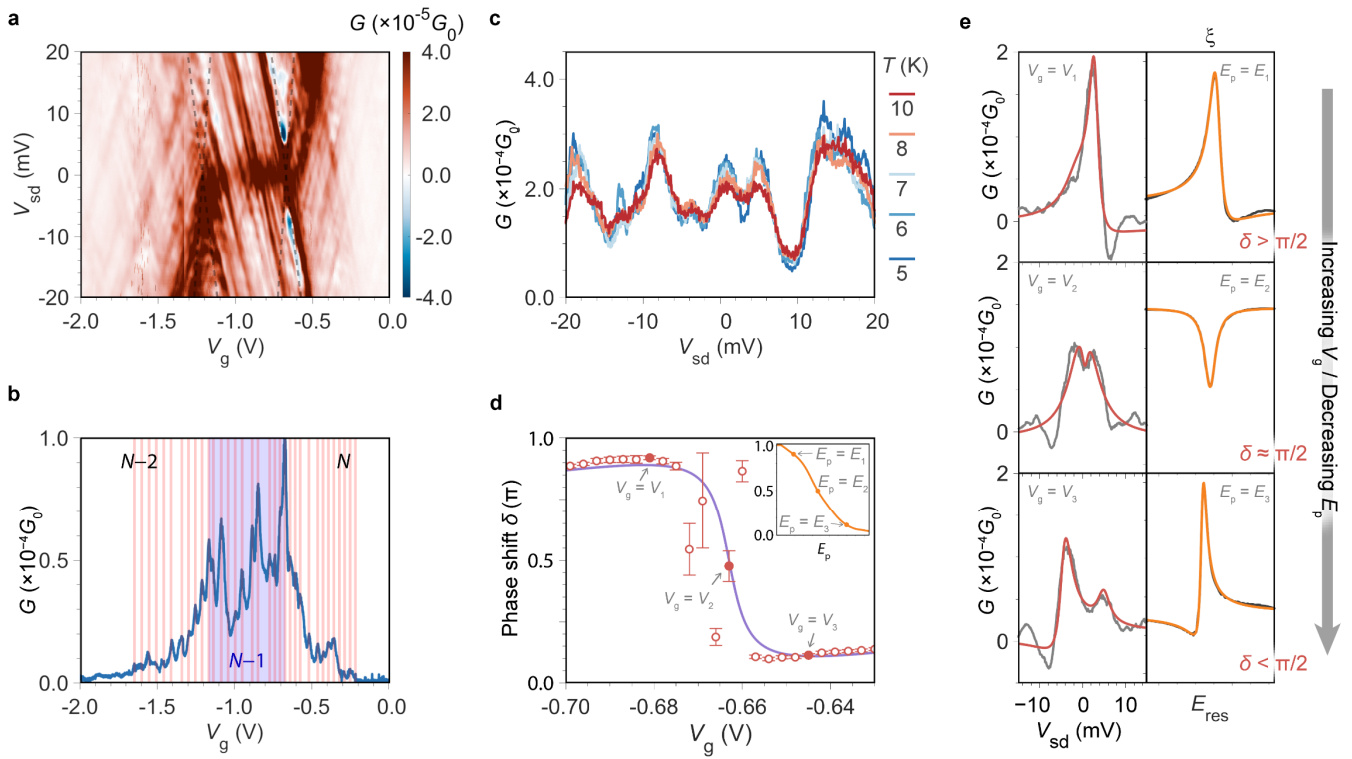

**Fig. 2 | Electron phase measurements for single molecule. a** Detailed conductance map, with dashed lines outlining molecular resonances. **b** $G$ vs $V_g$ plot at zero bias, with the fringes highlighted in red, and the regions with different charging of the molecular dot shaded in different colours, together with the molecular charge state. **c** Temperature dependence of $G$ vs $V_{sd}$ plot at $V_g = -1$ V. **d** Phase shift $\delta$ (red open dots) vs $V_g$, while scanning the voltage through resonance, taken from fits in **e**. The filled dots show three cases of panel **e**, that demonstrate the roughly $\pi$ shift upon scanning through $V_g = -0.663$ V, the inset panel shows the phase shift calculated for an optical model as a function of photon energy, $E_p$, also taken from **e**. **e** Individual $G$-$V_{sd}$ traces (grey) and fits (red) for different phase difference, $\delta$, conditions (left panels) at $V_1 = -0.681$ V (top), $V_2 = -0.663$ V (middle), $V_3 = -0.645$ V (bottom). The calculated $\xi$-$E_{Res}$ traces for the optical model interferometer in the right panels (black) with fits (orange).

A high-resolution conductance map (Fig. 2a) shows graphene FP resonances characterized by a fringe spacing of $\varepsilon$ = 4 ± 1 meV, which corresponds to a FP cavity length of ~0.9 µm (Supplementary Section 4.1),[11] in agreement with the KPFM data. The coherence is given by the visibility parameter $\eta = \frac{(G_{max} - G_{min})}{2\bar{G}}$, where $G_{max}$, $G_{min}$, and $\bar{G}$ are the maximum, minimum and mean conductance (Fig 2c). The local visibility (as mentioned above, it is $V_g$ dependent) at $V_g$ = –1.0 V is $\eta$ = 32% at 5 K and displays an exponential decrease with temperature. A fit to $\eta(T) = e^{-2\pi^2 k_B T/e_T}$ gives $\varepsilon_T = 7$ meV, which is slightly larger than the measured fringe spacing, indicating that thermal broadening of electron distributions in the leads is the primary coherence-loss mechanism,[24,25] but with a slightly faster decay that is likely the result of phonons or other inelastic processes. The characteristics of the graphene FP resonator thus allows interferometric detection up to 10 K, instead of mK temperatures.[6,24,26]

For phase detection it is required that the molecular and FP channels do not just coexist but interfere. Therefore, we study transport around the $N$–1/$N$ transition, at $V_g$~ –0.65 V, where the resonant tunneling through the molecular orbital (specifically the HOMO), intersects with FP resonances. We find evidence for interference in this region as the low-bias conductance takes the form of Fano resonances (Fig. 2e). Fano resonances are characteristic of a localised channel interacting with delocalised states in atomic, photonic, and nano-electronic systems.[27] Fano resonances have been described in single-molecule measurements, causing low conductance, and arise from intramolecular interference between differently localized orbitals, which are baked into the chemical structure.[28,29] In our case, these two channels are attributable to a single molecular orbital state (bound to the molecular structure) and the FP resonance extending over the graphene (Fig. 1e, left panels). The form of the Fano line shape depends on the phase difference, $\delta$, between the different paths, and therefore this can be extracted through fitting. The conductance is modelled by one Breit-Wigner (BW) resonance[30,31] term describing the FP transmission, and a second term for the Fano line-shape:

$$G = G_{FP} + G_{Fano} = A_{FP} \frac{\Gamma^2}{(\varepsilon - \varepsilon_{FP})^2 + \Gamma^2} + A_{Fano} \frac{[\tilde{\varepsilon} + \cot(\delta)]^2}{\tilde{\varepsilon}^2 + 1}$$

where $\Gamma$ is the FP resonance full width at half maximum, and $\varepsilon_{FP}$ its energy, with $\tilde{\varepsilon} = (\varepsilon - \varepsilon_{Mol})/(\Gamma_{Fano}/2)$ the dimensionless reduced detuning where $\varepsilon_{Mol}$ is the molecular resonance. There is good agreement with experimental data, and we find that the phase difference, $\delta$, remains > $\pi$/2 (Fig. 2e top) until the molecular and FP resonances both cross at zero bias and $V_g$ = –0.663 V, where the transition from $\varepsilon_{FP} < \varepsilon_{Mol}$ to $\varepsilon_{FP} > \varepsilon_{Mol}$ occurs (see Supplementary Fig. S4-3 for $\varepsilon_{FP}, \varepsilon_{Mol}$ vs $V_g$). At this point the two channels are simultaneously on resonance and we find $\delta \sim \pi$/2, leading to a pronounced avoided-level-crossing[32] and expected conductance dip at the centre of the resonance peak (Fig. 2e middle). The phase difference then abruptly shifts to $\delta$ > $\pi$/2 (Fig. 2e bottom), as indicated by the reflected Fano line shape, leading to a total shift of around $\pi$ (Fig. 2d) that is continuously tuneable through the gate voltage.

It is instructive to corroborate our model with a simulation of an equivalent optical structure.[32] The graphene electronic FP cavity is simulated as a vertical optical waveguide, while the molecule is represented by a coupled resonator (see Supplementary Fig. S5-1 for the modelled optical structure). The length of the waveguide adjusts the energies of the FP resonances ($E_{res} \propto 1/d$, where $d$ is the length), and the photon energy ($E_p$) can be scanned to bring the resonator in and out of resonance, as a simulation of $V_g$ on the molecular level. The curves of reflectance, $\xi$, vs $E_{res}$, and their fits to equation 2 indeed show the same behaviour of the $G$ vs $V_{sd}$ traces (Fig. 2e right panels), displaying Fano line-shapes, as in the electrical measurements, and the same fitted phase shift, $\delta$, when scanning through the resonance by varying $E_p$ (inset, Fig. 2d). The complete map of $\xi$ as a function of $E_{res}$ and $E_p$ also shows features comparable to the conductance maps, with the FP resonance peak transforming into a low-transmission dip and giving a fitted $\delta \approx \pi$/2 phase shift between the FP mode and Breit-Wigner mode, matching the results of our electronic experiment. This simulation confirms our graphene-molecule electronic interferometer works analogously to an optical FP interferometer and is a way to effectively probe the transmission phase shifts of electrons passing through the single molecule.

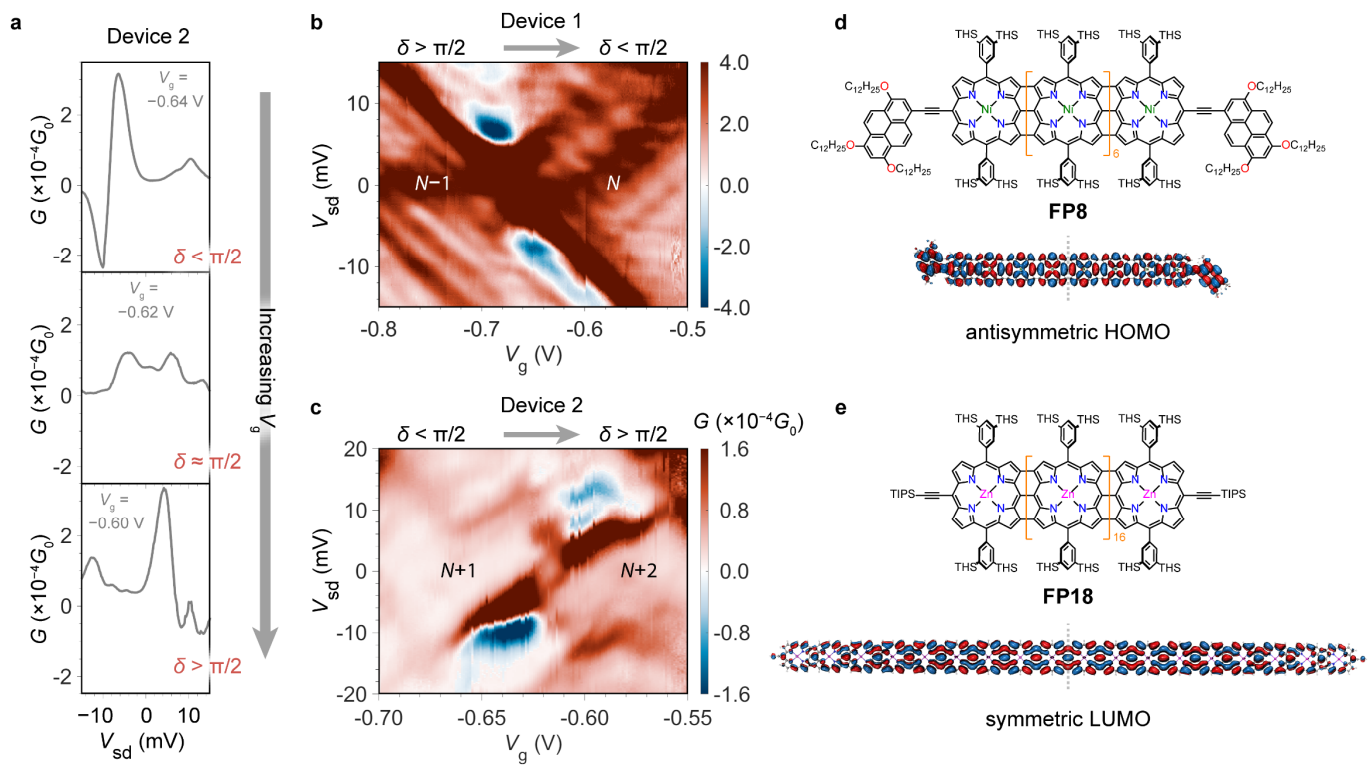

**Fig. 3 | Transmission phase shift and molecular orbital parity. a** Low bias $G$-$V_{sd}$ traces around the $N+1$ / $N+2$ transition of device 2 (using **FP18**) displaying evolution of Fano line shapes and transmission phase with $V_g$. **b-c** Comparative high-resolution conductance maps of Device 1 (**b**) and Device 2 (**c**), showing reversed phase shift behaviour. **d-e** Chemical structures of **FP8** (**d**) and **FP18** (**e**) and their antisymmetric HOMO and symmetric LUMO, respectively, that are the localized transport channels studied in **b** and **c**. **FP8** contains pyrene anchoring groups, whereas **FP18** is terminated by TIPS (triisopropylsilyl) groups. Both nanoribbons are made soluble through bis-3,5-(trihexylsilyl)phenyl groups, the difference in the metal ion (Ni(II) for **FP8**, Zn(II) for **FP18**) is to facilitate the chemical synthesis, as described in the text. THS: trihexylsilyl.

We used the same device structure to explore interferometric measurements through the longer porphyrin nanoribbon **FP18**, comprising 18 coupled porphyrin units, instead of 8. The metal centres in FP18 are Zn(II), rather than Ni(II), which reduces the HOMO-LUMO gap, and the effective mass of the charge carriers, prolonging the phase-coherence length. The nanogaps used to fabricate devices from **FP18** were of a similar size to those for **FP8**, so the much greater overlap with the graphene electrodes made anchor groups unnecessary for **FP18**, simplifying the synthesis. **FP18** is embedded in a graphene FP cavity from solution in the same manner, and the resulting conductance map shows the interference fringes of the electronic FP resonator (Supplementary Fig. S3-7). Shifting of the Fano resonances is observed with increasing $V_g$ (Fig. 3a), with the $\delta \sim \pi/2$ point at $V_g$ = –0.640 V and corresponding a conductance dip at zero bias (Fig. 3a middle, and Fig 3c). It is notable, though, that the transmission phase shift is now from $\delta < \pi/2$ to $\delta > \pi/2$. For device 1, the measurement was performed at $N$–1/$N$ transition of **FP8**, where the transport channel is the antisymmetric HOMO (Fig. 3d). For device 2 the transition was probably measured at $N+1$/$N+2$ transition of **FP18**, where the transport channel is the symmetric LUMO (Fig. 3e), although due to the absence of a large band gap in the conductance map, this is a tentative assignment. With the addition of an extra node in each subsequent FP resonance, transmission phase for this channel increases by $2\pi$ between resonances, whereas transmission phase increases by $\pi$ tuning through a Breit-Wigner resonance, explaining the $\Delta\delta = \Delta\phi_{Mol} - \Delta\phi_{FP} = |\pi|$ phase shifts observed for device 1 and device 2.[7] If the molecular orbital and FP are initially out of phase ($\delta = \pi$), then transition through the resonance point brings them in phase ($\delta = 0$), as for device 1. If the FP resonance and orbital are initially in phase $\delta = 0$, then the two channels are out of phase after sweeping through resonance ($\delta = \pi$), as is observed with device 2. If the parity of the Fabry-Perot resonances is known for device 1 and device 2, then the reversed phase behaviour could be linked to parity of the orbital involved, as different phase shifts ($\Delta\phi_{Mol}$) are a result of orbital symmetry.[33] For concrete assignments, moving towards more well-defined one-dimensional Fabry-Perot cavity geometries, thereby removing multimodal resonances, would be desirable. Smooth transitions of $\Delta\delta \sim |\pi|$ for both devices indicate that charging is not accompanied by orbital re-ordering (i.e., for device 1, the HOMO of the $N$ state and SOMO of $N$–1 state is of the same parity), which would result in an abrupt phase change (phase slip),[34] as observed in magnetoconductance measurements on quantum dots, or the lack of a phase shift. Regions with negative differential conductance can be found at finite $V_{sd}$, and the origin of this is not clear, but could be attributed to the out-of-equilibrium nature of the Fano resonances, asymmetric voltage drop across the device, or an interaction with an intruder continuum state under non-equilibrium conditions.[35] This might have potential impact on the accuracy of precise phase measurement, especially when the effect is strong (e.g. in device 2), but it is beyond the scope of this study.

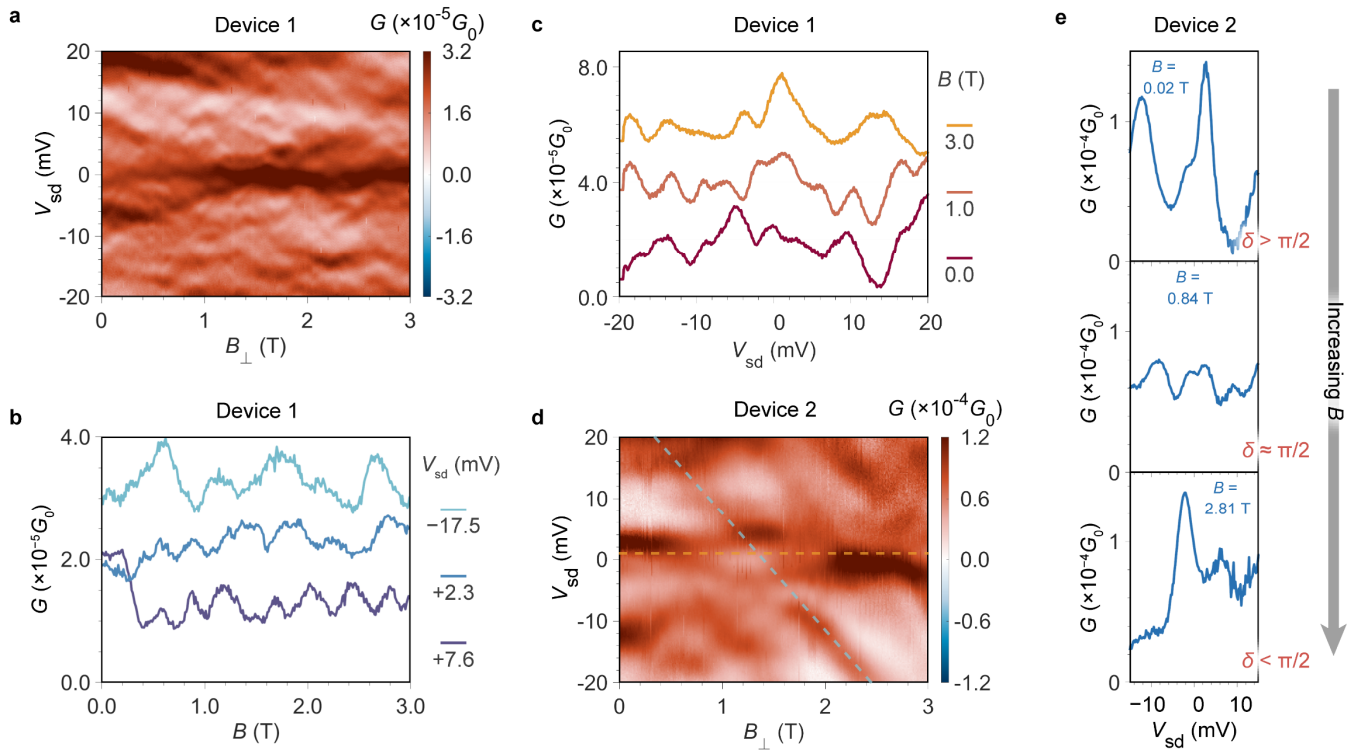

**Fig. 4 | Electron interferometry via electric and magnetic field tuning. a** $G$ map measured as a function of $V_{sd}$ and magnetic field ($B$) measured at $V_g$ = –1.00 V. The applied magnetic field is perpendicular (⊥) to transport plane (graphene). **b** $G$ plotted as a function of $B$ with $V_g$ = –1.00 V at fixed $V_{sd}$. The top curve ($V_{sd}$ = -17.5 mV) is offset +1.5 ×10$^{-5}$ $G_0$ for clarity. **c** $G$ plotted as a function of $V_{sd}$ with $V_g$ = –1.00 V at fixed $B$. The top curves ($B$ = 3.0 T) and middle curve ($B$ = 1.0 T) are offset +4.0 ×10$^{-5}$ $G_0$ and +2.0 ×10$^{-5}$ $G_0$ respectively for clarity. **d** $G$-$V_{sd}$-$B$ map of Device 2 measured at $V_g$ = –0.54 V, the orange and blue dotted lines indicate the molecular resonance and FP resonance, respectively. Note that the resonance at zero bias is the $N$+1 / $N$+2 transition at –0.64 V in Fig. 3, it is shifted here due to charge traps in the gate oxide. **e** individual $G$-$V_{sd}$ traces for Device 2 measured at different $V_g$ with fixed $B$ = 0 T (left panels) and at different $B$ with fixed $V_g$ = –0.54 V (right panels), showing Fano line shapes at different phase difference, δ, conditions.

Finally, we probe how magnetic field, $B$, alters the response of the interferometer. Measurements of conductance vs bias and $B$ (applied perpendicular to the graphene plane) for both devices shows a strong field-dependent response (Fig. 4). FP fringes shift linearly by > 10 meV/T, a much greater sensitivity than transport signals relating to the molecular resonance, e.g., the shift of a Coulomb peak due to the Zeeman effect is ( 1/2 $g_s\mu_B$ ~) 0.06 meV/T. Traces at fixed $V_{sd}$ yield magnetoconductance oscillations with periodicities of $\Delta B$ ~ 0.3 T (Fig. 4b), which can be related to the area enclosing electron trajectories through

$A = 2\pi h/e\Delta B$, which is of the order of 1 μm², in line with the dimensions of the FP cavity. For comparison, the areas of the nanoribbons are much smaller at ~ 8 nm² and ~ 16 nm² for **FP8** and **FP18** respectively (just considering the delocalized π-systems). The difference in coupling of the two channels to the magnetic field permit analogous experiments to the electrostatic tuning described above. Close to the $N+1$ / $N+2$ resonance for device 2 (Fig. 3c), there is a FP resonance, initially at $V_{sd}$ = 25 mV, that can be tuned by the magnetic field through the molecular resonance. This is functionally the same as the effect of increasing $V_g$ in Fig. 3, however it is a different FP resonance (higher in energy). The FP comes from high energy to low, accordingly, the evolution of the Fano line shapes as the two channels are brought into resonance and then detuned is opposite to Fig. 3a, the phase difference is initially $\delta > \pi/2$, and reaches $\delta \approx \pi/2$ (confirmed by the conductance dip at the anti-crossing) at $B$ = 0.84 T, with a subsequent drop in $\delta$ ($< \pi/2$) beyond this point (Fig. 4e). These results demonstrate additional capability of the interferometer design, with both electric and magnetic fields able to control of transmission electron phase through the devices.

Overall, these findings demonstrate a new approach to measuring transmission phase shifts at the nanometre scale without reliance on superconducting electrodes or magnetic field. Through the formation of an electronic Fabry-Pérot resonator with graphene nano-electrodes and coupling a single molecule, interfering channels can be tuned by external stimuli through their different electrostatic coupling to the gate potential. The interferometer offers a moderate visibility at a temperature that is two orders of magnitude higher than previous research on much large semiconductor nanostructures.[2,5] Future steps to create a better defined one-dimensional graphene FP cavity, and increasing the coherence length through encapsulation or mechanical exfoliation, will improve the visibility and remove multimodal resonances allowing for unambiguous assignment of molecular features. That said, the potential we have demonstrated for the detection and manipulation of transmission phase is of fundamental importance for both characterizing orbital and topological states in nanometre-scale objects, and would enable a parity readout mechanism that has been suggested[36] for quantum information processing on the scale of individual molecules and nanoribbons. More generally, experiments on optical interferometry offers the most precise measurements of length (for example, the LIGO experiment); future developments in electronic interferometry may offer more precise (sub-Γ) energy resolution in electrical measurements.

## Methods

*Substrate fabrication.* The substrates were fabricated using the following procedure. On a degenerately *n*-doped silicon wafer with a layer (300 nm thick) of thermally-grown silicon dioxide ($SiO_2$), a local gate electrode (3 μm wide) was defined by optical lithography with lift-off resist and electron-beam (e-beam) evaporation of titanium (5 nm thick) and platinum (15 nm thick). A layer (10 nm) of hafnium dioxide ($HfO_2$) was then deposited using atomic layer deposition (ALD). Next, source and drain contact electrodes separated by a 7 μm gap (the centre of the gap was aligned to the centre of the gate electrode, which means a 2 μm of horizontal distance between each electrode and gate electrode) were also defined by optical lithography with lift-off resist and electron-beam (e-beam) evaporation of titanium (5 nm thick) and platinum (45 nm thick).

*Graphene nanogaps.* A layer (600 nm) of poly(methyl methacrylate) (PMMA) (with a molecular weight of 495 kDa) was spin coated onto chemical vapour deposition (CVD)-grown graphene (purchased from Grolltex) on copper. The copper was then etched in aqueous ammonium persulfate (($NH_4)_2S_2O_8$) solution (3.6 g in 60 mL water) for 4 hours, after which the PMMA protected graphene was transferred 3 times to Milli-Q water and scooped up using the substrate. Air bubbles were further removed by partly submerging the sample in 2-propanol (IPA). The sample was dried overnight and baked at 180 °C for 1 h. The PMMA was then removed in hot acetone (50 °C) for 3 h.

The Z-shaped graphene tape with bow-tie shaped structure was patterned by e-beam lithography (EBL) with bi-layer lift-off resist (PMMA495 and PMMA950) and thermal evaporation of aluminium (50 nm thick). The Z-shaped graphene pattern was used so the inner graphene leads are coplanar with the bowtie structure (see Fig. 1b), reducing tension on the bowtie-shaped graphene, and maximising the stability of the junction. PMMA e-beam resist was used as it is positive resist and it can be transformed into smaller molecules after exposure, which make it easier to be completely removed than negative photoresist. Aluminium was then deposited onto exposed area as oxygen plasma resist, as aluminium can be completely removed by either acidic or basic aqueous solutions. By this method we reduce contamination from residual photoresist on graphene. The flatter configuration and cleaner surface might provide stronger molecule-electrode coupling by better molecule-graphene interfacing. After lift-off, the graphene on unexposed areas (which are not covered by aluminium) was etched with oxygen plasma. The aluminium was subsequently removed by aqueous sodium hydroxide (NaOH) solution (0.5 M; 1.0 g in 50 mL water). The sample was finally immersed in hot acetone (50 °C) overnight to remove any residual PMMA. The optical image and SEM images can be found in Supplementary Information Section 1.

Graphene nanogaps were prepared by feedback-controlled electroburning of the graphene bow-tie shape until the resistance of the tunnel junction is exceeded 1.3 GΩ ($10^{-7}$ $G_0$), as shown in Fig. S1-5. The empty nanogaps were characterised by measuring a current map as a function of bias voltage ($V_{sd}$) and gate voltage ($V_g$) at room temperature in order to exclude devices containing residual graphene quantum dots[18], only clean devices were selected for further measurement.

*Molecule junctions and measurements.* The solution of the porphyrin nanoribbon (1 μM in toluene) was drop-cast and allowed to dry in air. 124 devices for **FP8** and 257 devices for **FP18** were screened respectively. Only devices that showed clean current maps before molecule deposition were carried on further measurement. Thus, new signals appeared after molecule deposition can be attributed to transport through molecular junctions. Then, the chip contain molecular devices was connected to chip holder via wire bonding, loaded in Oxford Instruments 4 K PuckTester, and cooled down to cryogenic temperature for detailed measurements. All data shown in Figs. 1, 2, 3b and 4a-c are from **FP8** molecular device 1. All data shown in Figs. 3a, 3c and 4d-e are from **FP18** molecular device 2. Unless otherwise stated for temperature-dependant measurement, device 1 was measured at 4.2 K, while device 2 was measured at 2.8 K. The current maps and differential conductance maps can be found in Supplementary Information Section 3.

## Data Availability

All the data supporting the findings of this study are available within the Article, its Supplementary Information or from the corresponding authors upon request.

## Acknowledgements

The authors would like to acknowledge the use of the University of Oxford Advanced Research Computing (ARC) facility in carrying out part of this work (http://dx.doi.org/10.5281/zenodo.22558). The work was supported by the EPSRC (grants EP/X026876/1, EP/N017188/1 and EP/R029229/1), EIC-

pathfinder-4DNMR and EU-CoG-MMGNRs. JAM acknowledges funding from the Royal Academy of Engineering and a UKRI Future Leaders Fellowship, Grant No. MR/S032541/1. We thank Dr Xinya Bian and Dr Jacob L. Swett for help with the device substrates fabrication.

**Author Contributions**

The experiments were conceived by Z.C. and J.O.T with support from H. B., J.A.M., H.L.A. and L.B.; Z.C. performed the graphene patterning and fabrication of molecular devices and undertook the charge-transport measurements; J-R.D. synthesized and characterized the compounds under the supervision of H.L.A.; M. W. performed the optical simulation; N.F. performed the KPFM measurement. J.B. prepared the device substrates. Z. C. and J.O.T. analysed the data, and wrote the paper; all authors discussed the results and edited the manuscript.